\documentclass{natureHack_wFigsinplace}

\usepackage{amsmath}
\usepackage{amssymb}
\usepackage{hyperref}
\usepackage{capt-of}
\usepackage{multirow,booktabs}
\usepackage{url}
\usepackage{color,soul}  
\usepackage{ulem}
\usepackage{graphicx}

\newcommand{\icarus}{Icarus}
\newcommand{\mnras}{MNRAS}
\newcommand\aj{\ref@jnl{AJ}}

\title{The tumbling rotational state of 1I/`Oumuamua}

\author{Wesley C. Fraser$^{1\ast}$, 
Petr Pravec$^{2}$, 
Alan Fitzsimmons$^{1}$,
Pedro Lacerda$^{1}$,
Michele T. Bannister$^{1}$,
Colin Snodgrass$^{3}$,
Igor Smoli\'c$^{4}$
}

\begin{document}

\maketitle

\begin{affiliations}
\item Astrophysics Research Centre, Queen's University Belfast, Belfast, BT7 1NN, United Kingdom
\item Astronomical Institute, Academy of Sciences of the Czech Republic, Fri\v{c}ova 1, CZ-25165 Ond\v{r}ejov, Czech Republic
\item Planetary and Space Sciences, School of Physical Sciences, The Open University, Milton Keynes, MK7 6AA, UK
\item Scientific Computing Laboratory, Center for the Study of Complex Systems, Institute of Physics Belgrade, University of Belgrade,
Pregrevica 118, Zemun 11080, Belgrade, Serbia
\end{affiliations}

\begin{abstract}
The discovery\cite{Meech_2017} of 1I/2017 U1 (`Oumuamua) has provided the first glimpse of a planetesimal born in another planetary system. This interloper exhibits a variable colour within a range that is broadly consistent with local small bodies such as the P/D type asteroids, Jupiter Trojans, and dynamically excited Kuiper Belt Objects\cite{Ye_2017,Fitzsimmons_2017,Bannister_2017,Jewitt_2017,Bolin_2017,Masiero_2017}. 1I/`Oumuamua appears unusually elongated in shape, with an axial ratio exceeding 5:1\cite{Meech_2017,Jewitt_2017,Bannister_2017,Knight_2017}. Rotation period estimates are inconsistent and varied, with reported values between 6.9 and 8.3 hours\cite{Bannister_2017,Jewitt_2017,Feng_2017,Bolin_2017}. Here we analyse all available optical photometry reported to date. No single rotation period can explain the exhibited brightness variations. Rather, 1I/`Oumuamua appears to be in an excited rotational state undergoing Non-Principal Axis (NPA) rotation, or tumbling. A satisfactory solution has apparent lightcurve frequencies of 0.135 and 0.126 hr$^{-1}$ and implies a longest-to-shortest axis ratio of $\gtrsim5:1$, though the available data are insufficient to uniquely constrain the true frequencies and shape. Assuming a body that responds to NPA rotation in a similar manner to Solar System asteroids and comets, the timescale to damp 1I/`Oumuamua's tumbling is at least a billion years. 1I/`Oumuamua was likely set tumbling within its parent planetary system, and will remain tumbling well after it has left ours. 
\end{abstract}

\begin{figure}
\includegraphics[width=6in]{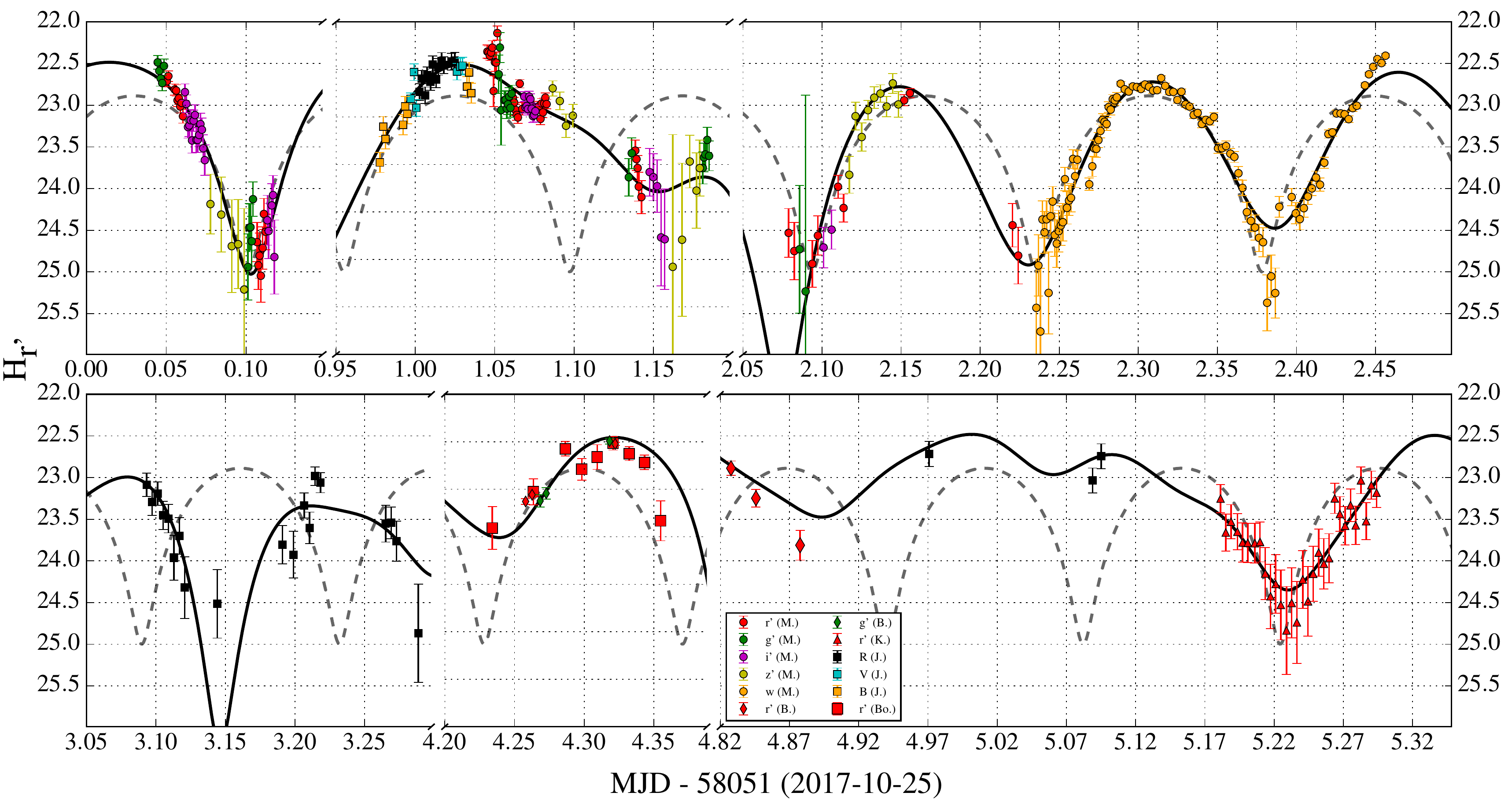}
\caption{The geometry-reduced and colour-corrected r'-band photometry of 1I/`Oumuamua cannot be well described by a model of simple rotation; the dashed line depicts the best nominal period of 6.831 hours. The tumbling model lightcurve (solid line), however, is an adequate representation (see Methods for comments on the quality of the model fit to the lightcurve minima). Data sources\cite{Bannister_2017,Jewitt_2017,Meech_2017,Knight_2017,Bolin_2017} including some reanalyzed data (see Methods) are indicated by common symbols and the first letter of their lead-author name. \label{fig:lightcurve}}
\end{figure}

Models of uniform rotation about a single spin axis moderately match 1I/`Oumuamua's observed brightness variations within a few nights\cite{Meech_2017,Bannister_2017,Jewitt_2017,Knight_2017,Bolin_2017}. These models are inadequate for the 6-night span of collated photometry we consider here (see Methods), with no single rotation period adequately matching the full set of data (see Figure~\ref{fig:lightcurve}). We find that models which consider linear increases or decreases in the spin period fare no better, producing equally inadequate matches. Similar conclusions have been arrived at by investigators using a different dataset\cite{Drahus_2017}.

A tumbling model\cite{Pravec_2005} with NPA rotation and free precession does provide an adequate description of the photometry (see Figure~\ref{fig:lightcurve}).  The relative sparseness of the data prevents determination of a unique set of frequencies; possible frequencies include 0.31, 0.26, 0.23, 0.16, 0.14, 0.12, 0.10, and 0.009~hr$^{-1}$. We have discounted values which are clearly commensurate with the Earth's rotation, although there is the possibility that one may be real. It is clear that tumbling provides a reasonable explanation for the peculiar brightness variations of 1I/`Oumuamua, which cannot be explained by simple single-axis rotation. Unfortunately, given the finite set of observations due to the limited observability of the object, it is unlikely that a unique solution will ever be determined for this object. 

The most complete lightcurve previously published\cite{Meech_2017} has an amplitude of $\sim2.5$ magnitudes, from which a minimum axial ratio of $a/c\sim$10:1 was suggested. Observing at non-zero phase angles $\alpha$, however, can enhance the apparent lightcurve amplitude by an amount dependent on the optical surface scattering properties of the body. An enhancement of up to $0.018$ magnitude/degree can occur for carbonaceous surfaces\cite{Gutierrez_2006}, and $\sim0.03$ magnitude/degree for S and similar type asteroids\cite{Zappala_1990}, due to the combined effects of optical scattering law, global shape and spin-pole inclination to the line of sight. As this early lightcurve was observed at $\alpha\simeq 20^\circ$, and our dataset spans $19<\alpha<24.5^{\circ}$, the true {\it conservative} lower limit to the axial ratio from observations is $a/c \gtrsim 5$. We caution, however, that the optical surface scattering properties of interstellar objects are unknown at present. 

NPA rotation of small asteroids with slow rotation periods is a well known phenomena\cite{Pravec_2005}. Tumbling can be brought about by collisions\cite{Henych_2013}, tidal torques in planetary close encounters\cite{Scheeres_2000}, cometary activity\cite{Samarasinha_2013}, or the YORP effect\cite{Vokrouhlicky_2007}. It is eventually damped by internal friction and stress-strain forces removing the excess of rotational energy above that of the basic rotational state around the body's principal axis with largest moment of inertia. The timescale to return to principal axis rotation depends on the body's internal rigidity and anelasticity, its initial rotation rate, density, size and shape\cite{Burns_1973,Sharma_2005,Breiter_2012}. As it is possible that 1I/`Oumuamua is either an icy comet-like body or more similar to organic-rich asteroids in the outer asteroid belt, we have estimated the timescale to return to principle axis rotation for both possibilities (see Methods). We find that a rigid, organic-rich body with the observed apparent elongation and size\cite{Meech_2017} will take $4\times10^{11}$ to $4\times10^{12}$~years to stop tumbling.  We note that recent theories\cite{Breiter_2012, Pravec_2014} suggest that the timescales may be a factor of 7--9 shorter than estimates that use the classical formula\cite{Burns_1973}.
For icy bodies, the damping timescale is an order of magnitude lower. A large fraction of Solar System asteroids smaller than $\sim200$ meters are tumblers, even among the very fast rotators that have tensile strength, e.g. 2000 WL107\cite{Pravec_2005} and 2008 TC3\cite{Scheirich_2010}.  Their tumbling suggests that they have a higher rigidity than larger and more weakly structured  asteroids, which results in long damping timescales.  1I/`Oumuamua may have a similarly prolonged damping timescale, which could be much longer than the age of the universe.

What induced the current tumbling of 1I/`Oumuamua?
As the YORP effect scales with incident stellar flux, and cometary activity only occurs within close proximity of a star, both should be negligible in interstellar space.  The space density of interstellar objects similar in size or larger than 1I/`Oumuamua is estimated\cite{Engelhardt_2017,Trilling_2017,Meech_2017,Jewitt_2017} as $n\sim0.1$ au$^{-3}$  and they should have a typical local standard of rest encounter velocity of $v\simeq25$ km/s\cite{Grav_2011}. The collisional lifetime for an interstellar object of effective radius $R$ is $\sim (n R^2 v)^{-1}\sim 10^{19}$ years. With a lack of other mechanisms available, it is clear that the tumbling of 1I/`Oumuamua probably commenced in its home planetary system.

\begin{figure}
\includegraphics[width=6in]{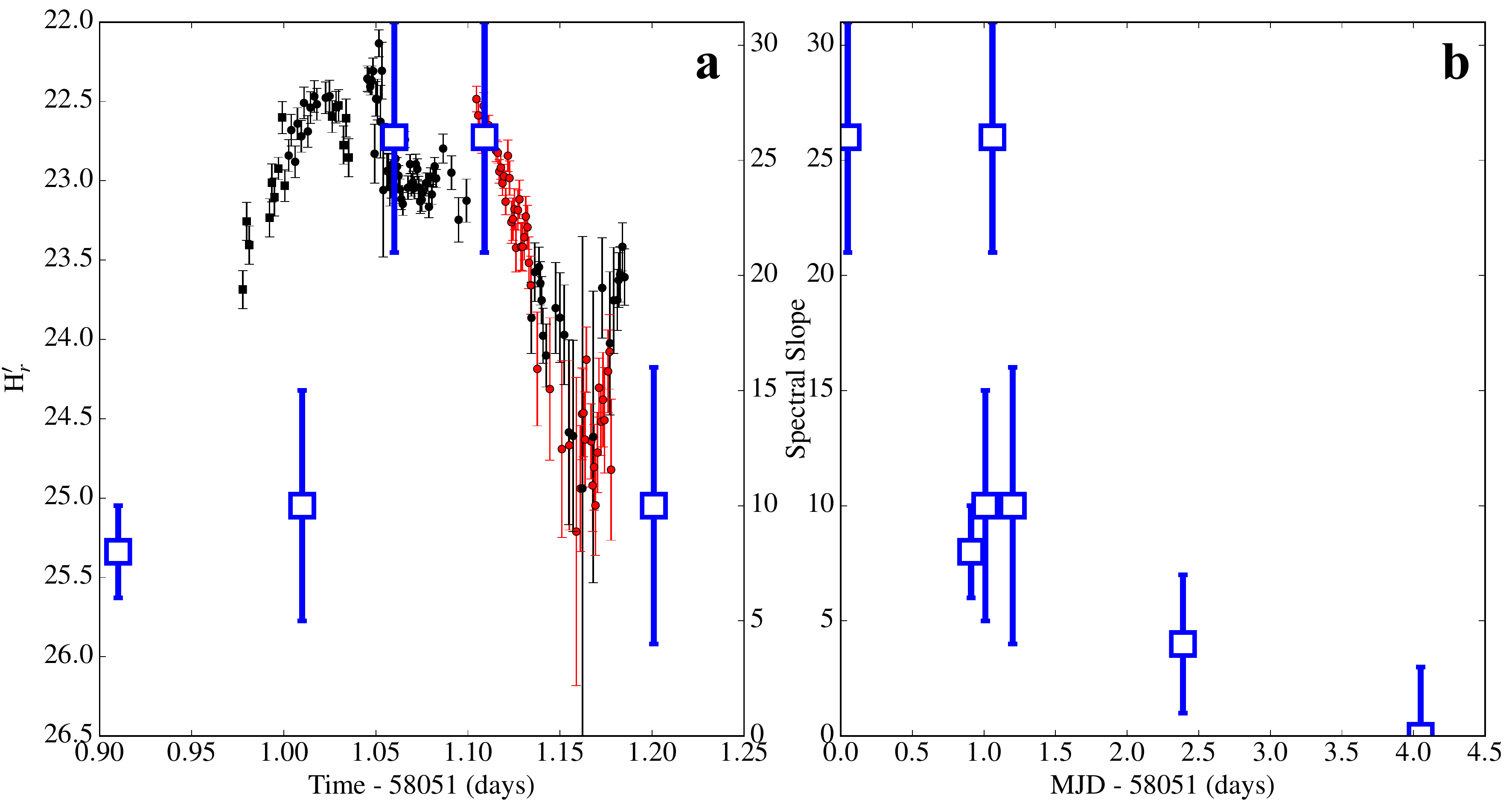}
\caption{\textbf{a:} Photometry of 1I/`Oumuamua from MJD 58051 (red) shifted to match the lightcurve peak observed during MJD 58052 (black). Colour measurements of 1I over this timespan are shown by the blue squares. All available colour measurements are presented as a function of time in panel \textbf{b}. Of the reported spectral slopes, only the reddest measurements correspond to peaks in the observed or modelled lightcurves, with the rest corresponding to lightcurve non-maxima. This argues for a body that is mainly a neutral reflector, with a red region, though available colour measurements are insufficient to determine if both of the long sides of 'Oumuamua are red, or just one. \label{fig:clc}}
\end{figure}

To estimate the shape of 1I, a detailed physical model of its tumbling will be needed.  While lightcurve inversion\cite{Pravec_2014} may be possible, such a technique makes a key assumption that the target possesses a uniform albedo, though to date 1I/'Oumuamua's albedo remains unknown. Instead, we consider colour, noting that colour may not correlate with albedo.  Within the six-day span available, colour variations have been detected for 1I/`Oumuamua, suggesting a compositionally varied surface. Reported optical spectral slopes (see Methods) span $0\lesssim S' \lesssim 25\%$/100 nm. The veracity of the detected colour variations is demonstrated by the consistency of the different overlapping photometric datasets, after correction to r' using the reported colours from each dataset. Large colour variations cannot explain the unusual lightcurve behaviour of our multi-band dataset, as even when considering only observations made in r', a single rotation period cannot be found. The observed colour variations appear to be correlated with the rotation of 1I/`Oumuamua (see Figure~\ref{fig:clc}). Though only two red ($S'\sim25\%$/100 nm) measurements have been reported, both fall after the brightest phases of the lightcurve observed in the first two nights of the photometry sequence. All other colour measurements correspond to lightcurve non-maxima and are neutral. The colour measurements imply that the body is largely a nearly-neutral reflector with spectral slope $S'\sim5\%$/100 nm, and has a large red region. 

'Oumuamua's colour variation is of a magnitude similar to the colour variations seen on some Kuiper Belt Objects\cite{Lacerda_2008,Fraser_2015}. It should be noted that the observed colours are roughly progressively more neutral with date observed (see Figure~2 panel b). It is tempting to attribute this to a trend of the surface of 1I/`Oumuamua progressively evolving to a more neutral colour with time, as is observed in some Jupiter Family Comets and Centaurs\cite{Jewitt_2015}. One of the red measurements, however, is bracketed by 3 neutral measurements taken over a span of a few hours on MJD 58052 (see Figure 2 panel b)  which is hard to explain through surface colour evolution. Rather, this colour sequence supports the idea of a spotted surface. Moreover, the idea of global colour evolution would require 1I/`Oumuamua's transition from red to neutral to occur entirely within the six-day duration of the observations, 42--48 days after perihelion, in an unlikely serendipity of timing. It appears 'Oumuamua's surface is inhomogeneously coloured.

\newpage

\begin{methods}
{\bf Optical photometry.} This was compiled from Bannister et al., Bolin et al., Jewitt et al., Knight et al., and Meech et al.\cite{Bannister_2017,Bolin_2017,Jewitt_2017,Meech_2017,Knight_2017} Where available, observations were corrected to the r' filter in the Sloan Digital Sky Survey (SDSS) photometric system \cite{Fukugita_1996}, using colours reported in those references, or using the temporally closest (g-r) colours\cite{Meech_2017, Bannister_2017}, assuming linear reflectance spectra through the optical range, an assumption that is consistent with reported spectra\cite{Ye_2017,Fitzsimmons_2017,Meech_2017}. The published and corrected photometry are tabulated in the Supplementary Table 1.

Where temporal overlap between photometry datasets was available, consistency within the measurement uncertainties was checked. Only one inconsistency was found, occurring  between the Gemini data from Bannister et al. and the photometry from Bolin et al. The Bolin data were reanalyzed using techniques similar to those made use of in Bannister et al. Specifically, median stacks of 5 images per stack were produced from the r' data. The Bolin data were all acquired with the telescope tracking at near the rate of 'Oumuamua, though moderate tracking errors were apparent from deviations away from linearity in the stellar shapes. This necessitated the use of an elongated kernel with which to measure star centroids. The on-image length and angle of those kernels was equal to the stellar trailing expected in each image. Stellar centroids were used for image alignment before stacking. Stacks were photometrically calibrated with in-image SDSS stellar catalogue sources with (g-r) colour within 0.4 magnitudes of Solar. Photometry of stars was measured using pill apertures, and of 'Oumuamua using a circular aperture with the Trailed Image Photometry in Python package\cite{Fraser_2016}. The aperture radii for stars and 'Oumuamua were 12 pixels, or roughly 2 FWHM, as any smaller resulted in divergent photometry, presumably as a result of the tracking errors. The resultant reanalyzed photometry was found to be fully consistent with the Bannister et al. Gemini measurements (see Figure~\ref{fig:lightcurve}). Original photometry, and colour and geometry corrected absolute magnitudes are available in digital format.

Spectral slopes (percent increase in reflectivity per 100~nm increase in wavelength normalized to 550~nm) were estimated from reported colours\cite{Bannister_2017,Jewitt_2017,Meech_2017} and spectra\cite{Fitzsimmons_2017,Ye_2017} assuming a linear spectrum through the optical range.

{\bf Lightcurve modelling.}
Simple rotation periods were searched for using phase dispersion minimization. The tumbling lightcurve was modelled following the technique of Pravec et al. (2005)\cite{Pravec_2005}: a 2-dimensional 2nd order Fourier series was used to model the observed brightness variations, and satisfactory solutions were searched for in the 2-dimensional frequency space.  A number of possible combinations of frequencies were found to fit the lightcurve data; one of those  providing a plausible fit is presented in Fig.~1.  
Note, however, that due to the limited coverage of the lightcurve with the available data (missing observations from the longitudes of Asia/Australia in particular) the solution is not ``anchored'' at uncovered times between the observational runs, allowing the fitted model to possibly under- or over-estimate the brightness variation at those times.  Without additional data breaking the commensurability of the observations with Earth's rotation, this problem cannot be solved. With the limited data, no Fourier series of order higher than 2 could be used.  While the 2nd-order Fourier series describes the lightcurve data well at most rotational phases, some lightcurve minima are not accurately modelled, with the fit either underestimating or overestimating the depth.  Additional data might enable use of a 3rd order Fourier series which would improve the model fit to the lightcurve minima.

{\bf Damping Timescales.}
To estimate the damping timescales $\tau_D$, we used the formulation used previously in the study of small Solar system bodies\cite{Burns_1973}:

$$\tau_D\simeq\frac{\mu Q}{\rho K_3^2 R^2 \omega^3}$$

\noindent
where $\mu$ is the rigidity of the object, $Q$ is the anelasticity or damping constant, $\rho$ is
the bulk density and $ \omega$ is the angular velocity of rotation. For 1I/`Oumuamua we
take $\omega=2.36\times10^{-4}\mbox{  s$^{-1}$}$ from our analysis. $K_3$ is a scaling coefficient that depends
on the oblateness of the body $p=(a-b)/a$ where $a$ and $b$ are the semi-major and semi-minor
axes of the body; $K_3\simeq 0.1\, p^2$. The lightcurve of Meech et al. along with an assumed
geometric albedo of $0.04$ implies projected radii $200\times20$ m ignoring amplitude-phase angle effects (see main text), hence we use this to estimate a maximum oblateness of $p\simeq0.9$ and $K_3\simeq 0.08$.

For an icy (or initially icy) comet-like body we assume $\rho\simeq 1000$ kg m$^{-3}$. The internal
rigidity is unknown, but we assume that $\mu\simeq 4 \times 10^9$ N m$^{-2}$ as assumed for previously
assumed for small bodies with weak strength\cite{Harris_1994}.
The mean radius is given by $R\simeq \sqrt{ab}\simeq 60$ m. Then  a range
of viable values of $100\leq Q \leq 1000$ in turn implies $4\times10^{10}\leq \tau_D\leq 4\times10^{11}$~years.
There is the possibility, however, that 1I/`Oumuamua is instead a 
rocky object similar to  C-type asteroids as found predominantly in the outer main-belt.
In this case we can assume $\rho \simeq 2000$ kg m$^{-3}$ and a rigidity a factor of 10 higher.
At the same time the geometric albedo could be higher at $\simeq 0.08$, implying a corresponding
reduction in $R$. This then predicts $4\times10^{11}\leq \tau_D\leq 4\times10^{12}$~years.

\subsection{Full Acknowledgements}

W.F., A.F., M.T.B., and P.L. acknowledge support from STFC grant ST/P0003094/1, and M.T.B. also from STFC grant ST/L000709/1.
The work by P.P. was supported by the Grant Agency of the Czech Republic, Grant 17-00774S.
C.S. is supported by a STFC Ernest Rutherford fellowship and grant ST/L004569/1.

\end{methods}




\begin{addendum}
 \item[Competing Interests] The authors declare that they have no
competing financial interests.
 \item[Correspondence] Correspondence and requests for materials
should be addressed to W. C. Fraser.~(email: wes.fraser@qub.ac.uk).
\item[Data Availability] All photometry used in this manuscript are tabulated in Supplementary Table 1.
\item[Author Contributions]
W.F. compiled the common filter dataset, re-reduced observations where necessary, and led the analysis and writing of the manuscript. P.P., P. L. and I.S. performed the lightcurve modelling and assisted with writing. A.F. calculated damping timescale estimates and assisted with paper writing. M.B. and C.S. assisted with interpretation of the lightcurve results and the paper writing.
\end{addendum}

\newpage

\end{document}